# Astro2020 Science White Paper

## The evolution of the cosmic molecular gas density

**Thematic Area: Galaxy Evolution**

**Principal Authors:** Fabian Walter (MPIA/NRAO), Chris Carilli (NRAO), Roberto Decarli (INAF OAS Bologna), Dominik Riechers (Cornell)

**Co-Authors:** Manuel Aravena (UDP), Franz Erik Bauer (PUC), Frank Bertoldi (AIfA Bonn), Alberto Bolatto (Maryland), Leindert Boogaard (Leiden), Rychard Bouwens (Leiden), Denis Burgarella (CNRS/CNES), Caitlin Casey (UT Austin), Asantha Cooray (UC Irvine), Paolo Cortes (JAO/NRAO), Pierre Cox (IAP), Emanuele Daddi (CEA), Jeremy Darling (Colorado), Bjorn Emonts (NRAO), Jorge Gonzalez Lopez (UDP), Jacqueline Hodge (Leiden), Hanae Inami (Hiroshima), Rob Ivison (ESO/Edinburgh), Ely Kovetz (BGU), Olivier Le Fèvre (LAM), Benjamin Magnelli (AIfA Bonn), Dan Marrone (Arizona), Eric Murphy (NRAO), Desika Narayanan (Florida), Mladen Novak (MPIA), Pascal Oesch (Geneva), Riccardo Pavesi (Cornell), Tanio Diaz Santos (UDP), Mark Sargent (Sussex), Douglas Scott (UBC), Nick Scoville (Caltech), Gordon Stacey (Cornell), Jeff Wagg (SKA), Paul van der Werf (Leiden), Bade Uzgil (NRAO/MPIA), Axel Weiss (MPIfR), Min Yun (UMass)

**Summary:** One of the last missing pieces in the puzzle of galaxy formation and evolution through cosmic history is a detailed picture of the role of the cold gas supply in the star-formation process. Cold gas is the fuel for star formation, and thus regulates the buildup of stellar mass, both through the amount of material present through a galaxy's gas mass fraction, and through the efficiency at which it is converted to stars. Over the last decade, important progress has been made in understanding the relative importance of these two factors along with the role of feedback, and the first measurements of the volume density of cold gas out to redshift 4, (the 'cold gas history of the Universe') has been obtained. To match the precision of measurements of the star formation and black-hole accretion histories over the coming decades, a two orders of magnitude improvement in molecular line survey speeds is required compared to what is possible with current facilities. Possible pathways towards such large gains include significant upgrades to current facilities like ALMA by 2030 (and beyond), and eventually the construction of a new generation of radio-to-millimeter wavelength facilities, such as the next generation Very Large Array (ngVLA) concept.

The Evolution of Galaxies since the Cosmic Dark Ages

Over the past two decades, the cosmic history of star formation has been quantified to great precision, due to the concerted efforts of numerous multi-frequency campaigns (see Madau & Dickinson 2014 for a review). These studies have shown that the comoving density of star formation rose by more than an order of magnitude from the dark ages to a broad peak around redshift ~ 2, when the Universe was only 3 Gyr old (Fig. 1, left). This peak constitutes the so-called 'epoch of galaxy assembly', when more than half of the stars in the current Universe were formed. Since this peak, the cosmic star formation has decreased steadily to its present day value, with a current star-formation rate density about an order of magnitude lower than at its peak. The time integral of this cosmic star-formation rate impressively matches (within a factor of 2) the stellar mass build-up that has been independently quantified throughout cosmic times (e.g. Muzzin et al. 2013). Attention is now turning to the other half of the cosmic baryon cycle, namely, the cold gas reservoirs that fuel star formation in galaxies.

Tracing the Cold Gas: the Fuel for Star Formation

Extensive studies of nearby galaxies show that star formation is fueled by cold gas, in particular, that in the molecular phase. In the local Universe, star formation is intimately linked to the presence of dense molecular clouds. It is in the densest parts of these clouds that molecular gas can cool and collapse, thus giving rise to new generations of stars. This molecular gas is mainly made up of molecular hydrogen ($H_2$). $H_2$ is almost impossible to observe in the physical conditions that are conducive to star formation (temperatures of <100K). Therefore, molecular structures are typically characterized with the help of tracer atoms or molecules in both our and external galaxies. In particular, it has been demonstrated that the carbon monoxide molecule (CO) is a good, albeit not perfect, tracer of the molecular gas content in normal galactic environments (see, e.g., Bolatto et al. 2013 for a review). CO remains a workhorse for investigating the molecular gas content in galaxies through cosmic time. Indeed, the vast majority of the few hundred molecular line detections reported so far at $z > 1$ have been in CO (e.g., Solomon & Vanden Bout 2005, Carilli & Walter 2013, Tacconi et al. 2018).

At sufficiently high redshift, millimeter wavelength interferometers like NOEMA and ALMA can observe the high rotational transitions of the CO molecule. These high-$J$ transitions are progressively sensitive to the warmer and denser molecular gas, e.g., star-forming or AGN regions. On the other hand, the lower-$J$ transitions, including the ground state CO ($J = 1-0$) transition, are only accessible to longer wavelength radio observatories (currently the Jansky VLA, and soon ALMA band 1). These lowest-$J$ transitions can directly trace the coldest phase of the molecular gas ($T \sim 10 - 100K$) and are the most robust tracers of the overall molecular gas mass, including any widespread or sub-thermally excited component. The conversion to a molecular gas mass $M_{H2}$ via the CO–to–$H_2$ conversion factor, $X_{CO}$ or $\alpha_{CO}$ remains uncertain. This conversion varies by up to a factor of 5 depending on the conditions of the gas in the ISM, including metallicity. Similar studies of the gas content have used the thermal dust continuum measurements to scale to gas masses (Magdis et al. 2011, 2012; Santini et al. 2014; Scoville et al. 2014, 2016, Genzel et al. 2015, Tacconi et al. 2018). However, this technique in many cases assumes that the dust-to-gas ratio is fixed across a wide range of galaxy types, redshifts, and metallicities, which can be a poor assumption in some cases (Rémy-Ruyer et al. 2014; Capak et al. 2015; Issa et al. 1990; Lisenfeld & Ferrara 1998; Draine et al. 2007; Bolatto et al. 2013; Berta et al. 2016, De Vis et al. 2019). Overall, all tracers, CO, dust luminosity, or the Rayleigh-Jeans flux density, require conversion factors and therefore have associated uncertainties. Hence continued studies using different tracers remain paramount. One unique advantage of CO (or any spectral line) is the additional kinematic information inherent in line observations, e.g., constraining $X_{CO}$ in high-$z$ galaxies is in principle possible via empirical measurements of galaxy dynamical masses, using e.g. the CO lines themselves.

At high redshift, the Karl G. Jansky VLA, GBT, and ATCA have detected CO(1-0) in a few dozen high-redshift starburst galaxies, using long (>10h) integrations on a few individual galaxies (e.g., Riechers et al. 2011; Ivison et al. 2011; Hodge et al. 2012; Emonts et al. 2014; Bolatto et al. 2013), or blind searches in deep fields with hundreds of hours on the VLA (Riechers et al. 2019; Pavesi et al 2018). Likewise, observations of the higher-$J$ CO lines with NOEMA and ALMA remain time intensive (e.g., Tacconi et al. 2018, Decarli et al. 2016b), with typical integration times of a few hours per target. Consequently, even after investing 1000s of hours on the most powerful facilities available today, only a few hundred high-redshift galaxies have been detected in CO line emission, and only tens of sources in the lowest-order transitions. These low numbers (cf. the extensive samples of available high-redshift galaxy spectra in the optical/NIR) remain the number one source of uncertainty in all molecular gas studies at high redshift.

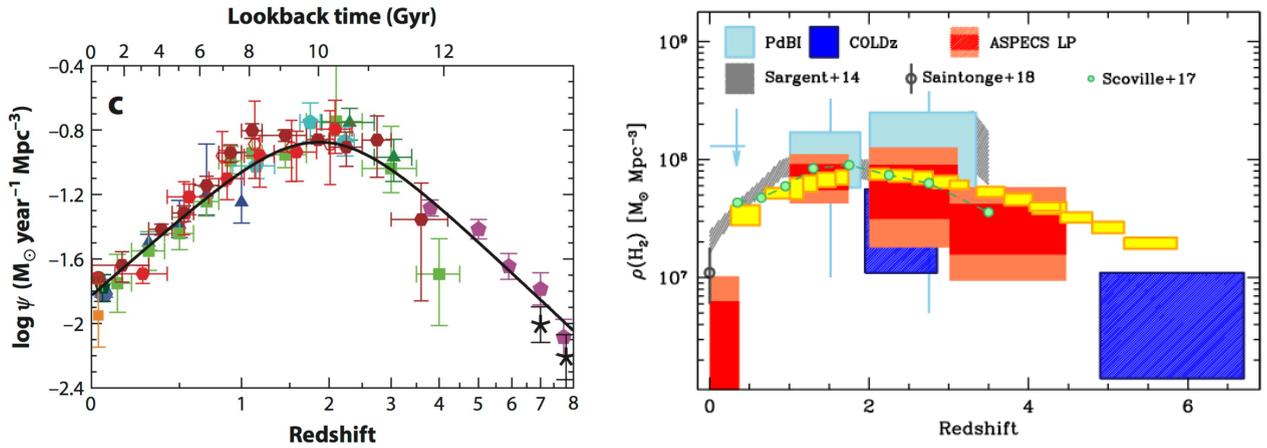

*Figure 1:* Left: Evolution of the cosmic star formation rate density is well constrained out to $z\sim6$, and shows a pronounced peak at $z\sim2$ ('the epoch of galaxy formation'). This plot is the result of decades of studies using multi-wavelength observations (taken from Madau & Dickinson 2014). Right: Corresponding molecular gas density ('the fuel for star formation') provides the currently missing piece in the cosmic galaxy build-up. First constraints from NOEMA, ALMA, and the VLA point at a similar decrease in molecular gas as seen in the star-formation rate diagram from $z\sim2$ to today. However, the current uncertainties are clearly substantial, and do not allow us to link the two plots on a galaxy-by-galaxy basis. An improvement of survey speed of order 100, which can only be afforded by a new generation of facilities (or vast extensions of current observatories), is essential to distinguish different mechanisms that can lead to the observed galaxy build-up. As an example, the yellow data points in the right panel show the precision that would be reached by a 1000-hour survey using the ngVLA concept.

The Evolution of the Cosmic Gas Density

Over the past decade, significant progress has been made to link the well-characterized cosmic star-formation rate density to the underlying molecular gas density. A full characterization of the fuel that eventually gives rise to the observed star formation is essential, since competing theories can explain the observed rise and fall of the cosmic star-formation history: e.g. it is conceivable that galaxies at the 'peak of galaxy assembly' form stars more efficiently out of their molecular gas content. Likewise, it is also plausible that the galaxies at cosmic noon had more gas, leading to increased star formation. Two independent and complementary approaches to address this fundamental question of galaxy evolution have been taken (see, e.g. Carilli & Walter 2013 for a review). (A) One is to directly measure the molecular gas content in a well-defined cosmic volume, through so-called 'frequency scans' using (sub-)millimeter and radio interferometers (ALMA, VLA, and NOEMA; e.g., Decarli et al. 2014; Walter et al. 2016; Pavesi et al. 2018). This method is still in its infancy, but it provides compelling evidence that the cosmic molecular gas density dropped from the 'peak of galaxy assembly' to today's value by an order of magnitude (Walter et al. 2014; Decarli et al. 2016a, 2019; Riechers et al. 2019), as shown in the right panel of Fig. 1. (B) The second approach is to measure the molecular gas fraction (typically defined as $M_{H2}/M_{stars}$) and star-formation efficiency (defined as $M_{H2}/SFR$) in galaxies at various redshifts. The local value of the molecular gas fraction is around 0.1, but at $z \sim 1$ to 2, this value rises by an order of magnitude, to unity, in high mass main sequence star forming galaxies (e.g., Daddi et al. 2010; Genzel et al. 2010, Tacconi et al. 2013, 2018). In other words, galaxies at high redshift are dominated (in terms of baryonic matter) by their molecular gas content, not their stellar mass. The decrease in gas fraction from $z \sim 2$ to 0 then parallels the same decline as seen in the cosmic star-formation rate density.

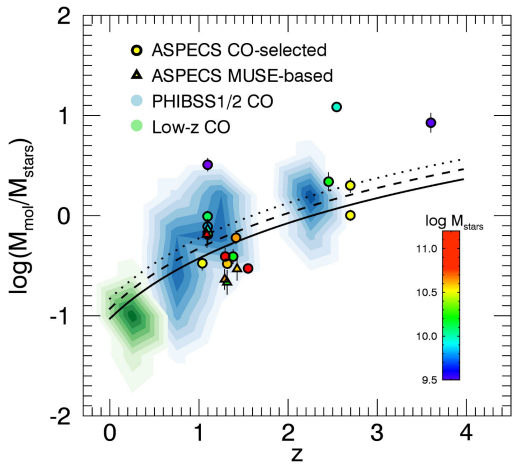

**Figure 2:** Gas mass fraction, defined as $M_{gas}/M_{stars}$ as a function of redshift / lookback time. At redshift z=0, this ratio is found to be of order 10%. At high redshift (z~2), the ratio is found to be unity or even higher. This implies that the baryons in galaxies at the 'peak of galaxy formation' start to be dominated by molecular gas, not stars. This has major implications for the star-formation process (pressure, densities). With current facilities, only the brightest and most massive main sequence galaxies can be detected. Taken from compilation in Tacconi et al. (2018) and Aravena et al. (in prep.).

A critical long-term goal is thus to have a precise measure of the dense gas history of the Universe, to connect it on a galaxy-by-galaxy basis to the history of cosmic star formation, i.e. to complement the available detailed measurements of the cosmic star-formation rate and stellar mass density. Such a measurement hinges on a two orders-of-magnitude increase in current survey speed, possibly afforded by an increase of collecting area, coupled with very wide bandwidth receivers to allow for efficient redshift searches, and possible multiplexing of beams to increase the instantaneous survey area. In Figure 1 (right) we show the current best published limits on the dense gas history of the Universe from the ALMA ASPECS,

VLA COLDz, and NOEMA HDF-N CO deep surveys (Walter et al. 2014; Decarli et al. 2016b, 2019; Riechers et al. 2019). We observe a general rise and fall with redshift, paralleling that of the star formation history of the Universe. However, current observations do not give insights into how the galaxies that contribute to the molecular gas density relate to those that define the cosmic density of star formation. Also, current constraints above redshift 3 are too loose to determine the cause for the increase of the star-formation rate at early times.

Open Questions in Galaxy Formation and Evolution

Important fundamental aspects of galaxy evolution are currently open, due to the current sparse samples and the related inability to connect the dense gas content to star formation activity for a wide range of galaxy properties. These limitations can only be addressed with a significant increase of collecting area and/or survey speed of current facilities operating in the millimeter-to-centimeter wavelength regime that match the capabilities of optical-to-infrared facilities that will ever more precisely detail the star formation rate density. Paramount questions include:

- *At the global level, to what degree is the increase of the star formation rate density from z=6 to 2 due to the increased availability of gas, rather than a change in the star formation efficiency?*
- *Which galaxies are the main contributors to the molecular gas density since the peak of galaxy formation, and within the first billion years of cosmic history?*
- *Are these the same galaxies that dominate the density of cosmic star formation?*
- *Does the rising gas fraction pertain only to the more massive galaxies? And does the gas fraction continue to rise at redshifts beyond $z > 2$?*

As demonstrated by recent investments of many hundreds to a few thousand hours following a suite of different strategies, these key questions cannot be answered with current facilities (in particular ALMA, VLA, and NOEMA), since at least tens of thousands of hours of observing time would be required (e.g., Tacconi et al. 2018; Riechers et al. 2019; Decarli et al. 2019). On the other hand, a 100-fold increase in survey speed in the coming decade(s) would make such investigations imminently feasible, positioning molecular studies in line with what is currently observed from high-redshift galaxies via direct starlight. Along with accurately constraining the history of the molecular gas density (e.g. yellow data in Fig. 1, right), large samples of molecular gas detections in the early Universe open up numerous critical avenues in galaxy formation studies, including: (i) detailed demographic studies of the molecular content as a function of galaxy properties, such as stellar mass and star formation rate; (ii) study of large-scale structure via 2-point correlation and power spectra, for the CO emitting population themselves, and cross-correlation with galaxies selected at other wavelengths, to understand the bias and halo masses inherent in molecular-selected samples; and (iii) studies of the environmental impact, such as local galaxy density or the cold circumgalactic medium, on the molecular gas content of galaxies. The resulting line luminosity functions would also provide a critical anchoring point for large-scale line intensity mapping surveys (e.g., Kovetz et al., Astro 2020 White Paper).

Such molecular gas surveys in the distant Universe will be performed in well-surveyed legacy fields, where there is already substantial ancillary data available, including optical/NIR spectroscopic redshifts (soon including measurements from JWST). With large numbers of spectroscopic redshifts, stacking to measure the median molecular gas reservoir in main sequence galaxies would become possible as a function of redshift, environment, or other physical factors.

The Future Path Forward

Decarli et al. (2018) use mock observations covering the CO transitions from $2 < z < 9$, to compute the flux distribution of CO emitters from state-of-the-art semi-analytical models of galaxy formation and evolution that are coupled to radiative transfer calculations and chemical abundances from photodissociation region models (Popping et al. 2012, 2014, 2016). These estimates are in general agreement with observational constraints on the CO luminosity functions at high redshift derived from NOEMA HDF-N, ALMA ASPECS, and VLA COLDz surveys (Walter et al. 2014; Decarli et al. 2016a; Riechers et al. 2019; Decarli et al. 2019). Based on these predictions, the expected number of CO detections for future facilities with one to two orders of magnitude increased sensitivity and survey speed (possible realizations could be an extension of ALMA beyond the ALMA2030 roadmap, or the concept study of a next-generation VLA [ngVLA]) will exceed 100 per pointing (compared to only one to a few in present surveys), or 3000 in a ~1000 hr survey over a region that is 100 times larger than the Hubble UDF.

This factor of 100 increase in volume probed will result in orders of magnitude improvement in measurements of the cold gas history of the Universe (see Figure 1, right) and the evolution of gas fractions compared to the largest surveys possible with current facilities. This will elevate them to comparable levels of precision as anticipated for synergetic measurements of the star formation, stellar mass, and black hole accretion histories from the upcoming generation of optical/infrared facilities, such as *JWST*, Euclid, WFIRST, US-ELT/E-ELT. Several of these facilities will also provide the spectroscopy necessary for stacking in the molecular line surveys far beyond the nominal survey limits, possibly complemented by the atomic line diagnostics provided by the proposed far-IR/sub-mm Origins Space Telescope. Together, these will also provide complementary key physical probes, such as the history of black hole growth and the rise of metals across the whole electromagnetic spectrum (e.g., Pope et al., Astro 2020 White Paper).

Deep fields at optical through radio wavelengths have shown the power of deriving key physical quantities of a large number of galaxies in the high redshift Universe. Large samples allow for detailed demographic studies of, e.g., the relative contribution to the cosmic star formation rate of galaxies with different stellar masses or star formation rates, and the large-scale clustering properties of these systems to infer halo masses (e.g., Elbaz et al. 2007, Magnelli et al. 2011). This is currently not possible for molecular gas studies. For instance, current investigations show clearly the order-of-magnitude rise in the gas-to-stellar mass ratio from $z = 0$ to $z \sim 2$ for main-sequence galaxies (Figure 2), indicating a fundamental change in their properties out to the epoch of galaxy assembly. However, a two orders of magnitude increase in sample sizes in low-$J$ CO measurements is indispensable to perform demographic studies, such as breakdowns by star-formation rate or stellar mass, or to investigate the role of feedback. This can be achieved by significant investments in telescope resources, such as considerable extensions of ALMA (more antennas, broader bandwidths, multi-beam receivers), or the ngVLA concept study.

**The evolution of the cosmic molecular gas density** 7